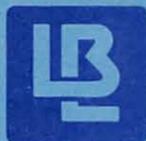 **Lawrence Berkeley Laboratory**
UNIVERSITY OF CALIFORNIA

Submitted to Physics Letters B

TRANSVERSE MOMENTUM ANALYSIS OF COLLECTIVE MOTION
IN RELATIVISTIC NUCLEAR COLLISIONS

P. Danielewicz and G. Odyniec

October 1984

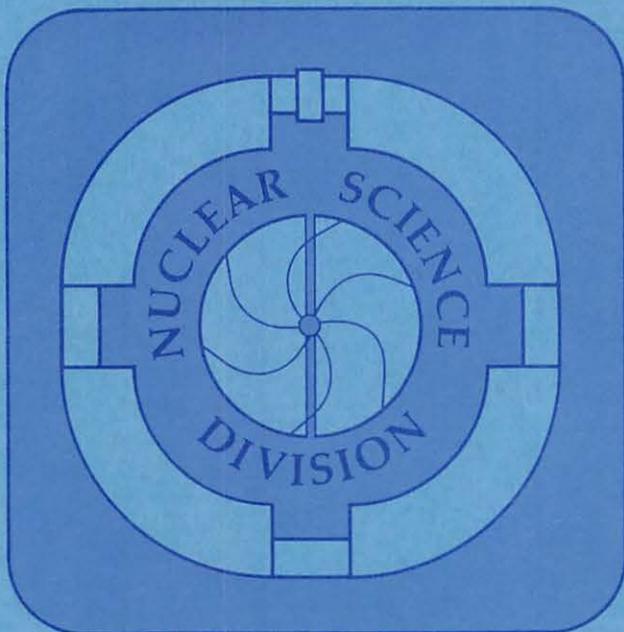

Transverse Momentum Analysis of Collective Motion

in Relativistic Nuclear Collisions*

P. Danielewicz[†] and G. Odyniec

Nuclear Science Division
Lawrence Berkeley Laboratory
University of California
Berkeley, CA 94720

Abstract

Novel transverse-momentum technique is used to analyse charged-particle exclusive data for collective motion in the Ar+KCl reaction at 1.8 GeV/nucl. Previous analysis of this reaction, employing the standard sphericity tensor, revealed no significant effect. In the present analysis, collective effects are observed, and they are substantially stronger than in the Cugnon cascade model, but weaker than in the hydrodynamic model.

* This work was supported by the Director, Office of Energy Research, Division of Nuclear Physics of the Office of High Energy and Nuclear Physics of the U.S. Department of Energy under Contract DE-AC03-76SF00098.

[†]On leave of absence from Institute of Theoretical Physics, Warsaw University, Warsaw, Poland.

Much experimental and theoretical attention in the study of high-energy heavy-ion collisions has been devoted to the possible existence of collective motion, following decompression of highly excited nuclear matter. Evidence has been claimed from two-particle correlations [1] and from sphericity analysis [2]. The extraction of collective flow parameters from data and the comparison with theoretical predictions, such as those of the cascade [3] and hydrodynamic [4] models is made difficult by the existence of statistical fluctuations. To recognize this, it is easy to see that a spherical momentum distribution, populated at random by a finite number of particles, will yield a reaction plane, an elongated sphericity tensor, and a finite collective flow angle. The magnitude of the uninteresting statistical effects will depend on the multiplicity of particles.

In this letter we isolate collective motion in the Ar+KCl(1.8 GeV/nucl) reaction [5], with a novel transverse-momentum analysis method. Transverse momentum (see also refs. [6]) is selected to avoid any possible effects due to nuclear transparency and corona [7] effects that would be manifested primarily in the longitudinal momenta. We determine the reaction plane in the collisions, and show how to estimate its uncertainty. We show how to remove finite-multiplicity distortions from events rotated to the reaction plane, and present the collective effects as the distribution of average transverse momenta projected onto the reaction plane, as a function of rapidity. By this means we are able to demonstrate collective motion in a reaction for which the sphericity method was inconclusive [5]. We complement the results with a new evaluation of the parameters of the sphericity matrix.

The semi-exclusive data of the near-symmetric Ar+KCl(1.8 GeV/nucl) reaction come from central-trigger measurements in the streamer chamber at the Bevalac. Details of the experiment have been reported previously [5]. The analysed 495 collision events were processed for protons, deuterons, tritons, $\pi^+$, and $\pi^-$ from the reaction. The central-trigger cross-section of 180 mb corresponds in the geometric picture to a cutoff in the impact parameter at $b < 2.4$ fm.

We define a vector constructed from the transverse momenta $p_\nu^\perp$ of detected particles:

$$\underset{\sim}{Q} = \sum_{\nu=1}^{M} \omega_\nu \underset{\sim}{p}_\nu^\perp \tag{1}$$

where $\nu$ is a particle index and $\omega_\nu$ is a weight. We choose $\omega_\nu = 0$ for pions. For the baryons we choose $\omega_\nu = 1$ for $y_\nu > y_c + \delta$, $\omega_\nu = -1$ for $y_\nu < y_c - \delta$, and $\omega_\nu = 0$ otherwise. For symmetric collisions it is natural to choose $y_c$ equal to the value of the overall c.m. system, $y_c = y_B/2 = 0.87$. The quantity $\delta$ is inserted to remove particles from midrapidity which do not contribute to the determination of the reaction plane but do contribute unwanted fluctuations. For this reaction we take $\delta = 0.3$ which excludes 35% of detected nuclear fragments. The direction of the vector $\underset{\sim}{Q}$ can finally be used to estimate the reaction plane in an event, and its magnitude to determine average momentum transfer in the reaction.

To estimate the accuracy of the procedure we divided randomly each event into two, and compared the reaction planes extracted from the two sub-events. The azimuthal correlation between the constructed vectors $\underset{\sim}{Q}_I$ and $\underset{\sim}{Q}_{II}$ is shown in fig. 1a. The fact that the distribution is not flat testifies to the determination of the reaction plane. To verify that this result is not due to inefficiencies in the streamer chamber acceptance, we performed a similar test using Monte-Carlo events generated by mixing particles from events within the
same multiplicity range. The resulting distribution is shown in fig. 1b. It
is completely flat. Arguing with the central-limit theorem and small-angle
expansion one can deduce that the distribution of the observed $\underset{\sim}{Q}$ with respect
to the reaction plane should be more narrow than that of fig. 1a by a factor
of 2. (A $\sqrt{2}$ factor in the width reduction comes from the increase in
multiplicity, and a $\sqrt{2}$ from the change from a deviation between two sampling
vectors, to a deviation from plane.) The optimal choice of $\delta$ in the
definition of $\underset{\sim}{Q}$ can be made to minimize the width of the distribution
in fig. 1a. To verify that the result is not dominated by a few particles, we
removed from each event the four particles with the highest transverse momenta
in $\underset{\sim}{Q}$. The correlation was slightly diminished, but remained within the error
bars of fig. 1a.

We turn to a discussion of the magnitude of $\underset{\sim}{Q}$. If $\underset{\sim}{Q}$ were just a sum of
randomly oriented momenta, then we should have the average $\overline{Q^2} = \overline{\Sigma p^{\perp 2}}$.
However, from the data we get

$$\overline{Q^2 - \Sigma p^{\perp 2}} = \sum_{\mu \neq \nu} \overline{(\omega_\mu \underset{\sim}{p}_\mu^\perp)(\omega_\nu \underset{\sim}{p}_\nu^\perp)} = 11.4 - 6.7 = 4.7 \pm 0.5 \ (\text{GeV/c})^2. \qquad (2)$$

To the extent that correlations, other than that stemming from existence of the
reaction plane (initial state of the collision), are weak,

$$\overline{Q^2 - \Sigma p^{\perp 2}} = \sum_{\mu \neq \nu} \overline{(\omega_\mu p_\mu^x)}\overline{(\omega_\nu p_\nu^x)}, \qquad (3)$$

with the averages at the r.h.s. of (3) taken in the coordinate system
associated with the reaction plane, and x denoting a vector component in the
reaction plane. With a fragment mass $a_\nu$, and the total mass $A = \Sigma a$, we
estimate average momenta per nucleon in the reaction plane from

$$\overline{\omega p^X/a} \simeq \left( \overline{(Q^2 - \Sigma p^{\perp 2})} / \overline{(A - \Sigma a^2)} \right)^{1/2} = 95 \pm 5 \text{ Mev/c,} \tag{4}$$

and the average transverse–momentum transfer in the reaction plane, $\overline{Q^X}$, with

$$\overline{Q^X} \simeq \overline{A} \cdot \overline{\omega p^X/a} = 2.17 \pm 0.11 \text{ GeV/c} \tag{5}$$

(for practical purposes this is square root of the numerical value of eq. (2)). Finally, for total transverse momentum in the forward region, $y > y_c + \delta$, in terms of nuclear charges, $\overline{P_f^X}$, we estimate

$$\overline{P_f^X} \simeq \overline{Z_f} \cdot \overline{\omega p^X/a} = 1.04 \pm 0.05 \text{ GeV/c.} \tag{6}$$

Utilizing (4), the experimental $\overline{p^{\perp 2}}$, and multiplicities, we attempted to simulate the data with two Gaussian sources associated with the reaction plane in the transverse–momentum space. The simulation reproduces nontrivial $p^{\perp}$ data–averages, and also distributions, like $dN/dQ^2$. Carrying with the sources the procedure as with data for fig. 1a, we get the dashed line in fig. 1a. The solid line in fig. 1a indicates the azimuthal angle distribution of $Q$ from the sources with respect to the reaction plane. The distribution is broad, $(\overline{\phi^2})^{1/2} \simeq 56^0$. For the subsequent analysis it is important whether $\overline{\cos\phi}$ is significantly larger than zero as compared with unity, and we get $\overline{\cos\phi} \simeq 0.65$.

We now proceed to establish the average transverse momentum per nucleon in the reaction plane as a function of rapidity $\overline{p^X/a}(y)$. We start with what might seem most natural [8], rotating events to a common reaction plane, for each event from $Q$, and evaluating the in–plane averages $\overline{p^{X'}/a}(y)$. The results are shown in fig. 2a. Since momenta are not projected on the true reaction plane, but on an estimated one, we put a prime on x. In fig. 2b we show results from the same procedure with the Monte–Carlo events that lack a dynamic effect in the reaction plane. The figures look similar because of the finite-multiplicity fluctuations [8]. Let us examine the distortion of momenta
in fig. 2, exhibited in the apparent collective effect for Monte-Carlo events.
As we project momenta on the reaction plane from $\underset{\sim}{Q}$, we evaluate

$$p_\nu^{x'} = \frac{p_\nu^\perp \cdot \underset{\sim}{Q}}{Q} = \frac{p_\nu^\perp \cdot \sum_\mu \omega_\mu p_\mu^\perp}{|\sum_\mu \omega_\mu p_\mu^\perp|} = \frac{\omega_\nu p_\nu^{\perp 2} + p_\nu^\perp \cdot \sum_{\mu \neq \nu} \omega_\mu p_\mu^\perp}{|\sum_\mu \omega_\mu p_\mu^\perp|}, \tag{7}$$

and for Monte-Carlo events

$$\overline{p^{x'}}(y) \sim \frac{\overline{\omega p^{\perp 2}}}{\sqrt{M} \, p^\perp} = \frac{\overline{\omega p^\perp}}{\sqrt{M}} \sim \frac{530 \text{ MeV/c}}{4.5} \sim 100 \text{ MeV/c.} \tag{8}$$

Here M stands for the number of particles contributing to $\underset{\sim}{Q}$, and we insert into
(8) values appriopriate for the reaction. The distortion $(\sim 1/\sqrt{M})$ occurs
because we project a particle momentum on itself. This is more general, and
the distortion would occur if the reaction plane were estimated from sphericity
matrix [2,8]. Once we relate a particle to a construct in which a particle has
been used, we probe a correlation of a particle with itself.

To remove the distortion of momenta, we determine the reaction plane for
each particle separately from the remaining particles in an event

$$p_\nu^{x'} = p_\nu^\perp \cdot \underset{\sim}{Q}_\nu / Q_\nu, \qquad \underset{\sim}{Q}_\nu = \sum_{\mu \neq \nu} \omega_\mu p_\mu^\perp. \tag{9}$$

The reevaluated $\overline{p^{x'}}/a(y)$ are shown in figs. 3a,b, for data and Monte-Carlo
events, respectively, and the distinction is now clear. Figs. 3c,d, show the
average differential, per unit rapidity, transverse momentum deposition in the
estimated reaction plane, $\overline{dP^{x'}}/dy$, in terms of nuclear charges. The variable,
integrated over a rapidity interval, measures total transverse momentum
deposited in nuclear charges in an interval. As the particle momenta are not
projected onto the true reaction plane, the average momenta get reduced:

$\overline{p^{x'}}(y) = \overline{p^x}(y) \cdot \overline{\cos\phi}$, where $\phi$ is the azimuthal angle deviation of the estimated plane from the true one. Normalizing momenta with the total observed momentum (5), l.h.s. scales in figs. 3a,c, we find $\overline{Q^{x'}}/\overline{Q^x} = \overline{\cos\phi} = 0.64 \pm 0.06$, in an agreement with the simulated distribution of $\underset{\sim}{Q}$ in fig. 1a.

With the result (6) and fig. 3c, we estimate total transverse momentum transfer between hemispheres ($y \lessgtr y_B/2$) in terms of nuclear charges of p, d, t from the reaction, at $2.3 \pm 0.2$ GeV/c. Assuming a corresponding amount of transfer in neutrons, would bring the total transfer in the reaction to $\gtrsim 4.9$ GeV/c. Here we comment on Coulomb effects. An estimate, from eq. (13.1) of ref. [9], shows that Coulomb repulsion cannot contribute to the reaction–plane transverse–momenta of forward– and backward–going nuclear fragments, more than ~10 MeV/c per particle. For pions we find from data at $|y - y_c| > \delta : \overline{\omega p^x_{\pi^-}} \simeq 12 \pm 8$ MeV/c, and $\overline{\omega p^x_{\pi^+}} \simeq 4 \pm 8$ MeV/c. A comparison (later) with the cascade model, excluding shadowing, singles out the decompression of excited nuclear matter as responsible for the observed collective motion in the reaction plane. Compared with general features of the reaction, the effect is moderate. Thus, cf. (4), r.m.s. transverse momenta are $\overline{(p^{12}/a)}^{1/2} \simeq 525$ MeV/c, and in one transverse direction $(1/\sqrt{2})\overline{(p^{12}/a)}^{1/2} \simeq 370$ MeV/c. Further, the particle distribution is much elongated in the beam direction, as will be emphasized by the variables from sphericity matrix, that we proceed to compute.

The vector $\underset{\sim}{S}^{\perp z} = \Sigma_\nu \underset{\sim}{p}^\perp_\nu p^z_\nu/(2m_\nu)$ is of the form (1), and may be subjected to the moment analysis as $\underset{\sim}{Q}$ in eqs. (2–5). Accordingly, we estimate a crucial element of the per–particle sphericity matrix, $s^{ij} = \overline{p^i p^j/(2m)}$, $\overline{S^{ij}} = \Sigma_\nu \overline{s^{ij}_\nu}$, from the formula analogous to (4)

$$\overline{s^{xz}} \simeq \left( (\overline{(S^{1z})^2} - \Sigma \overline{(s^{1z})^2}) / \overline{(M(M-1))} \right)^{1/2} = 31 \pm 2 \text{ MeV.} \tag{10}$$

Further, we evaluate the matrix elements $\overline{s^{zz}}$, $\overline{s^{x'x'}}$, $\overline{s^{y'y'}}$, and transform the latter two into $\overline{s^{xx}}$ and $\overline{s^{yy}}$. In an analogy with $\overline{p^x}(y)$, the transformation now involves $\overline{\cos^2\phi}$, that we estimate from simulated distribution in fig. 1a. Upon diagonalization $\widehat{\overline{s}} = \widehat{\text{diag}}(f_1, f_2, f_3)$, we find for the flow angle $\theta = 10.2 \pm 0.5^0$, and for the eigenvalue ratios $r_{31} = f_3/f_1 = 3.35 \pm 0.14$, $r_{32} = 3.09 \pm 0.12$, and $r_{21} = 1.09 \pm 0.08$. The index 2 is affixed here to the axis out of the reaction plane, and indices 1 and 3 to the axes shorter and longer in the reaction plane, respectively.

We now confront the findings with the theoretic models. From fig. 4 of ref. [7], with the result of $^{40}\text{Ca}+^{40}\text{Ca}(0.4 \text{ GeV/nucl})$ at $b = 2$ fm ideal-fluid [4] calculation, we read off transverse momenta per nucleon in the forward and backward rapidity regions $\overline{\omega p^x}/a \sim 200$ MeV/c. With the hydrodynamic scaling [10] this would correspond to $\overline{\omega p^x}/a \sim 400$ MeV/c at $E_{1ab} = 1.8$ GeV/nucl. Though particle production should soften the rise of momenta with energy, it is clear that the momenta from fluid dynamics at 1.8 GeV/nucl would exceed few times those observed experimentally. For the eigenvalue ratio and flow angle that are geometric and thus scaling invariant, the fluid dynamic model yields at relevant impact parameters [7], $r_{31} \simeq (2 - 4)$ and $\theta \simeq (30 - 60)^0$, with the angle in large excess to that observed experimentally.

Carrying calculations of the Ar+KCl(1.8 GeV/nucl) reaction at $b < 2.4$ fm, with the Cugnon [3] intranuclear code, we find for the nucleon momenta in the reaction plane $\overline{\omega p^x} = 22 \pm 2$ MeV/c, at $|y - y_B/2| > 0.3$. The average total transverse momentum of protons in the forward hemisphere is $\overline{P_f^x} = 0.23 \pm 0.06$ GeV/c, and the flow parameters are $\theta \simeq 2.6^0$, $r_{31} \simeq 3.7$, and $r_{32} \simeq 3.7$, and $r_{21} \simeq 1.0$. Since there may be some uncertainty in the impact parameters of the data, we quote also cascade results at fixed $b = 3$ fm: $\overline{\omega p}^X = 32 \pm 3$ MeV/c, $\overline{P_f^X} = 0.27 \pm 0.05$ GeV/c, $\theta \simeq 3.2^0$, $r_{31} \simeq 4.6$, $r_{32} \simeq 4.6$, $r_{21} \simeq 1.0$. We conclude that the cascade model underestimates transverse momenta of the data by a factor of the order of four. The sampling distribution, as in fig. 1a, is nearly flat for the cascade model. Estimating with (4) we are able to detect the minute momenta in the reaction plane, getting $\overline{\omega p}^X \simeq 28 \pm 7$ MeV/c, and $\overline{\omega p}^X \simeq 31 \pm 9$ MeV/c, at $b < 2.4$ fm, and $b = 3$ fm, respectively. For $\overline{s^{XZ}}$, we estimate with (10), $\overline{s^{XZ}} \simeq 9 \pm 2$ MeV, and $\overline{s^{XZ}} = 10 \pm 3$ MeV, while elements evaluated with the known reaction plane are $\overline{s^{XZ}} = 8 \pm 1$ MeV, and $\overline{s^{XZ}} = 12 \pm 1$ Mev, respectively.

The current results do not include corrections for the overall transverse momentum conservation. The corrections may be important when the moment analysis (3,4,10) is separately applied to the forward or backward hemispheres, for weaker dynamic effects than in the data.

To conclude, we have successfully isolated the collective motion effects in the transverse momenta from the Ar+KCl(1.8 GeV/nucl) reaction. The apparent ease with which one gets a handle on the motion with the new method of analysis, stems from the fact that the method explores the 2-particle correlation induced by existence of the reaction plane, amplified by summation over many particles. By contrast the standard sphericity method explores the single-particle [8] aspect of many particle distribution associated with the reaction plane.

Acknowledgements

We gratefully acknowledge the help from H. Pugh. We thank for the help and encouragement from all the other members of the LBL-GSI Streamer Chamber Group, in particular from J. W. Harris, R. Renfordt, and R. Stock. We want to to thank M. Gyulassy for comments and stimulating discussions. This work was supported by the Director, Office of Energy Research, Division of Nuclear Physics of the Office of High Energy and Nuclear Physics of the U.S. Department of Energy under Contract DE-AC03-76SF00098.

Figure Captions

Fig. 1. Azimuthal angle distribution of vectors $Q_I$ and $Q_{II}$ from subevents with respect to each other: (a) for data, (b) for Monte-Carlo events. Dashed line is for the simulation with two Gaussian sources. Solid line represents azimuthal angle distribution of vector $Q$ from the sources with respect to the reaction plane (this is normalized to the height with the sampling distribution).

Fig. 2. Average in-plane transverse momentum per nucleon as a function of rapidity, from rotation of events to the reaction plane determined by $Q$: (a) for data, (b) for Monte-Carlo events.

Fig. 3. (a,b) Average momentum per nucleon in the estimated reaction plane $\overline{p^{x'}}/a(y)$, upon removal of finite-multiplicity distortions, for data and Monte-Carlo events, respectively. (c,d) Differential, per unit rapidity, transverse momentum deposition in the estimated reaction plane in terms of nuclear charges $\overline{dP^{x'}}/dy$, for data and Monte-Carlo events, respectively. L.h.s. scales in (a) and (c) yield respective estimated average momenta per nucleon $\overline{p^{x}}/a(y)$, and deposition $\overline{dP^{x}}/dy$, in the true reaction plane.

XCG 8410-13380

Fig. 1

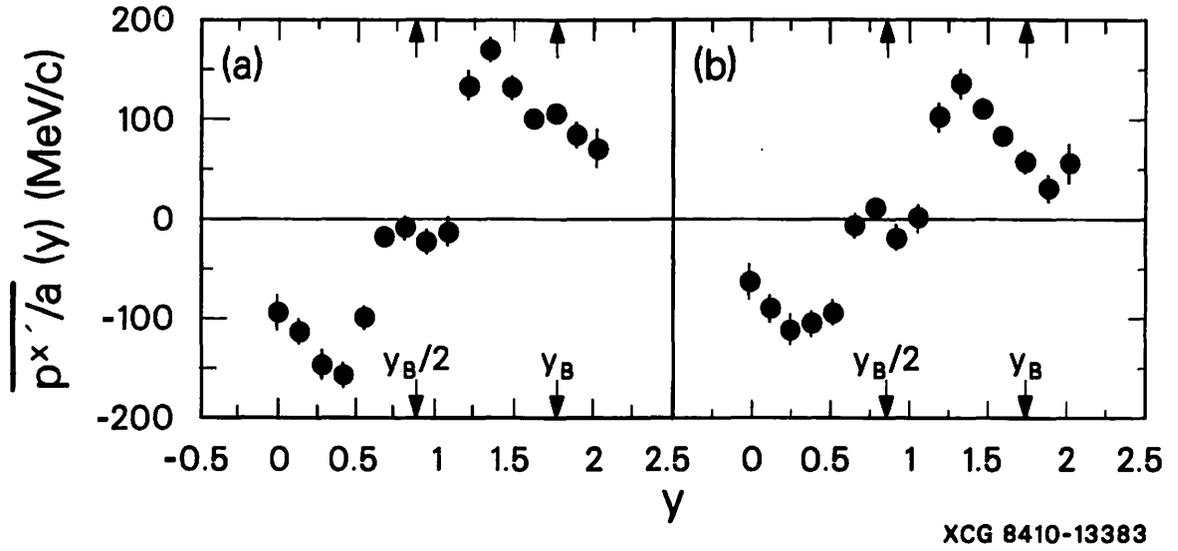

XCG 8410-13383

Fig. 2

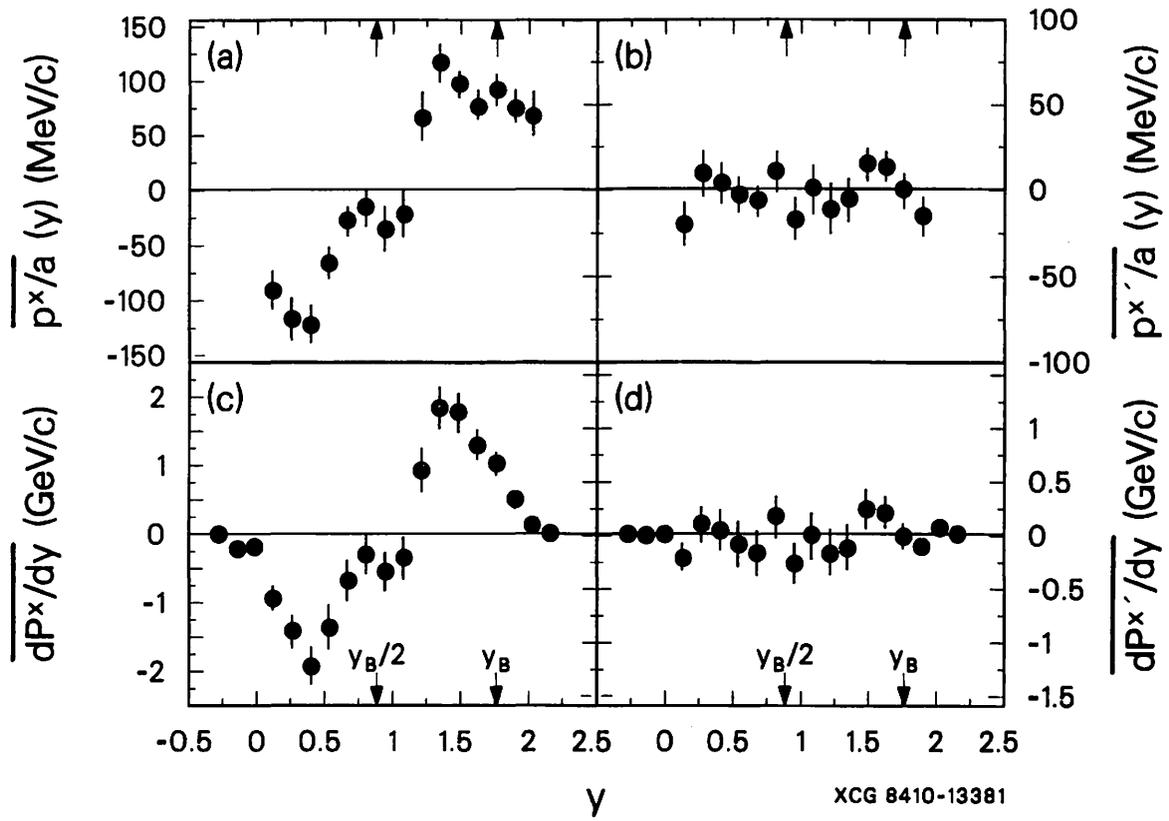

Fig. 3

XCG 8410-13381